\documentclass[a4paper,11pt]{extarticle}

\usepackage{multicol}

\usepackage[numbers,sort&compress]{natbib}

\usepackage{graphicx}
\usepackage{wrapfig}
\usepackage{cancel}
\usepackage{subfigure}
\usepackage{enumerate}
\usepackage{amssymb,amsmath}
\usepackage{epstopdf}
\usepackage{epsfig}
\usepackage{textcomp}
\usepackage{sidecap}
\usepackage{floatrow}
\usepackage{enumitem}
\usepackage{changepage}   
\usepackage{enumitem}
\usepackage{mathrsfs}
\usepackage{amsfonts}

\usepackage[a4paper, total={6.5in, 9.5in}]{geometry}

\usepackage{titlesec}
\titleformat{\section}[block]
  {\fontsize{15}{15}\bfseries\sffamily}
  {\thesection}
  {1em}
  {}
\titleformat{\subsection}[block]
  {\fontsize{12}{15}\bfseries\sffamily}
  {\thesubsection}
  {1em}
  {}

\newcommand{\be}{\begin{equation}}
\newcommand{\ee}{\end{equation}}
\newcommand{\ba}{\begin{eqnarray}}
\newcommand{\ea}{\end{eqnarray}}

\newcommand{\modifa}[1]{\textcolor{Black}{#1}} 

\begingroup\lccode`~=`_
\lowercase{\endgroup\def~}#1{_{\scriptscriptstyle#1}}
\AtBeginDocument{\mathcode`_="8000 \catcode`_=12 }

\usepackage[affil-it]{authblk} 
\usepackage{etoolbox}
\usepackage{lmodern}
\makeatletter
\patchcmd{\@maketitle}{\LARGE \@title}{\fontsize{22}{19.2}\selectfont\@title}{}{}
\makeatother

\let\OLDthebibliography\thebibliography
\renewcommand\thebibliography[1]{
  \OLDthebibliography{#1}
  \setlength{\parskip}{0pt}
  \setlength{\itemsep}{0pt plus 0.2ex}
}

\usepackage[usenames,dvipsnames,svgnames,table]{xcolor}

\title{Neural-Parareal: Dynamically Training Neural Operators as Coarse Solvers for Time-Parallelisation of Fusion MHD Simulations}

\author[ ]{S.J.P.Pamela$^1$, N.Carey$^{1}$, J.Brandstetter$^{2,3}$, R.Akers$^1$, L.Zanisi$^{1}$, J.Buchanan$^{1}$, V.Gopakumar$^{1}$, M.Hoelzl$^{4}$, G.Huijsmans$^{5,6}$, K.Pentland$^{1}$, T.James$^{1}$, G.Antonucci$^{1}$ and the JOREK Team$^{7}$}

\affil[.]{EUROfusion Consortium, JET, Culham Science Centre, Abingdon, OX14 3DB, UK.}
\affil[1]{CCFE, Culham Science Centre, Abingdon, Oxon, OX14 3DB, UK.}
\affil[2]{ELLIS Unit, Linz, LIT AI Lab, Institute for Machine Learning, Johannes Kepler University, Linz, Austria.}
\affil[3]{NXAI GmbH, Austria.}
\affil[4]{Max-Planck Institute for Plasma Physics, 85748 Garching, Germany.}
\affil[5]{CEA, IRFM, F-13108 Saint-Paul-lez-Durance, France.}
\affil[6]{Eindhoven University of Technology, 5612 AZ Eindhoven, The Netherlands.}
\affil[7]{See author list of \textnormal{M Hoelzl et al., Nuclear Fusion 61, 065001 (2021)}.}
\affil[.]{Corresponding author: Stanislas Pamela, stanislas.pamela@hotmail.com}

\date{\texttt{ }}

\begin{document}
\maketitle


\section*{Abstract}

The fusion research facility ITER is currently being assembled to demonstrate that fusion can be used for industrial energy production, while several other programmes across the world are also moving forward, such as EU-DEMO, CFETR, SPARC and STEP. The high engineering complexity of a tokamak makes it an extremely challenging device to optimise, and test-based optimisation would be too slow and too costly. Instead, digital design and optimisation must be favored, which requires strongly-coupled suites of High-Performance Computing calculations. In this context, having surrogate models to provide quick estimates with uncertainty quantification is essential to explore and optimise new design options. Furthermore, these surrogates can in turn be used to accelerate simulations in the first place. This is the case of Parareal, a time-parallelisation method that can speed-up large HPC simulations, where the coarse-solver can be replaced by a surrogate. A novel framework, Neural-Parareal, is developed to integrate the training of neural operators dynamically as more data becomes available. For a given input-parameter domain, as more simulations are being run with Parareal, the large amount of data generated by the algorithm is used to train new surrogate models to be used as coarse-solvers for future Parareal simulations, leading to progressively more accurate coarse-solvers, and thus higher speed-up. It is found that such neural network surrogates can be much more effective than traditional coarse-solver in providing a speed-up with Parareal. This study is a demonstration of the convergence of HPC and AI which simply has to become common practice in the world of digital engineering design.


\section{Introduction}

\subsection{Motivation}

Solving non-linear systems of partial differential equations is a field of research that has applications in a wide range of scientific and engineering problems. In the aerospace and automotive industries, in weather and climate predictions, in fusion energy research, countless numerical solvers are being used routinely to predict the evolution of complex physical systems. Conventional PDE solvers are constantly being developed as part of scientific research (MOOSE, MFEM, Firedrake, OpenFoam, JOREK) \cite{MOOSE,MFEMweb,MFEM_MHD2021,Firedrake,FiredrakeMHD2021,OpenFoam,JOREK,JOREK_overview,MOOSE} as well as industrial tools (ANSYS, ABAQUS, SIEMENS) \cite{ANSYS,ABAQUS,SIEMENS}. Modern PDE solvers are typically parallelised on the spatial domain they address, but in the case of fully implicit solvers, which have strong numerical stability advantages, this typically results in large matrix inversions with preconditioners. These typically do not scale well on large High-Performance Computing (HPC) systems with GPU accelerators due to memory limits and bandwidth. Methods to further parallelise conventional solvers have been explored in recent decades including, among others, parallel-in-time methods such as Parareal \cite{Lions-Parareal-2001} and deep learning methods such as neural operators \cite{Z.Li-FNO-2020,Brandstetter2022,Jiang2023}. While these emerging methods often lack the precision of the underlying conventional solvers they exploit, their efficacy and practical relevance is highly dependent on the use-case of interest. Their value lies in providing fast approximations of the detailed computation, which is of interest for wider integrated digital engineering tools and digital twins \cite{Chakraborty2021}. 

In fusion research, the design (and design optimisation) of new tokamak and stellerator devices requires a wide range of HPC calculations, to be integrated in a coupled workflow, that may require several steps to converge to a final solution. Some of these individual components needed for fusion power plant designs are themselves integrated workflows comprising of several HPC codes. For example, the blankets around a tokamak plasma, which will be used to breed tritium from the fusion-born neutrons and extract their energy into a cooling system, require neutronics calculations with codes like OpenMC or MCNP \cite{OpenMC,Romano2015-OpenMC,MCNP}, coupled to fluid mechanics \cite{Brooks2022} or even liquid-metal Magnetohydrodynamics (MHD) \cite{Mistrangelo2024}. However, the material and mechanical properties of these blankets are also strongly dependent on the heat-fluxes that result from plasma turbulence at the plasma edge \cite{Mason2019,Schwander2024}, which eventually requires integrated simulations from a plasma flight simulator like JINTRAC \cite{Romanelli2014}. This couples key characteristics of the plasma dynamics, such as the Grad-Shafranov equilibrium, turbulent pressure transport fluxes, deposition of various heating and fueling systems, MHD stability limits, and Scrape-Off Layer kinetic simulations. Breeding blankets and plasma dynamics are just two examples of the complex coupled system required to obtain a fully consistent digital design or digital twin of tokamak devices. Having the ability to accelerate some of these components, even to obtain an initial guess, can enable engineers to quickly explore a wider range of configurations to optimise the design of future machines.

The work proposed here combines two independent methods to provide a novel approach of accelerating conventional solver approximations. Namely, using neural operators as the coarse solvers required by the Parareal method \cite{Lions-Parareal-2001}. The convergence of High-Performance-Computing and Artificial Intelligence is illustrated by this approach, where the training of the neural operator is bootstrapped into the large data-production feature of the Parareal method. To demonstrate this approach, a generic fusion application is chosen, using a set of MHD equations in toroidal geometry, where filamentary blob structures are evolved inside a 2D slab domain. The outcome of this study is threefold: 1. the speed-up provided by the Parareal algorithm is increased by improving the accuracy of the coarse-solver, 2. a fast, accurate coarse-solver is obtained which can be used as surrogate inside other workflows to accelerate estimates of calculations, and 3. the accuracy of the coarse-solver can be defined by how fast the Parareal simulations converge, providing an uncertainty quantification of the surrogate.

\subsection{Current Research on Parareal and Neural Operators}

The current research on parallel-in-time methods is a wide field of science and the Parareal method \cite{Lions-Parareal-2001} is only one of its branches. The particularity of the Parareal method is that it is relatively straight-forward to understand and implement, with a non-intrusive implementation for the numerical tool, although in practice it requires the developer to understand in detail the i/o of the code in question. In this work, the code chosen for the demonstration uses Finite-Element Methods (FEM), which makes this aspect of the Parareal implementation more intrusive, as will be explained in detail in further sections. Applications of Parareal methods have been achieved in various fields of numerical studies, including Molecular Dynamics \cite{Legoll2022}, fluid dynamics \cite{Guilherme2021}, geodynamics \cite{Samuel2012}, as well as fusion \cite{Samaddar2019}.

The Parareal approach, although abstract, is relatively simple. A good introduction to Parareal can be found in \cite{Pentland2023}. It consists of splitting a time-domain into multiple \textit{time-windows}, and evolving each window concurrently. A first rapid estimate is done across all time-windows with a so-called \textit{coarse-solver}, which should have a negligible execution time compared to the full simulation code (often called the \textit{fine-solver}), which is run for each time-window in parallel, starting from the estimate provided by the coarse solver. The same procedure is repeated at each Parareal \textit{cycle}, where the initial-value for each time-window is calculated using a \textit{predictor-corrector} scheme, which combines the result of the previous time-window's coarse solution and the previous time-window's coarse and fine solutions from the previous cycle.

\begin{figure}[ht!]
  \centering
	\subfigure[ ]{\includegraphics[width=16.0cm]{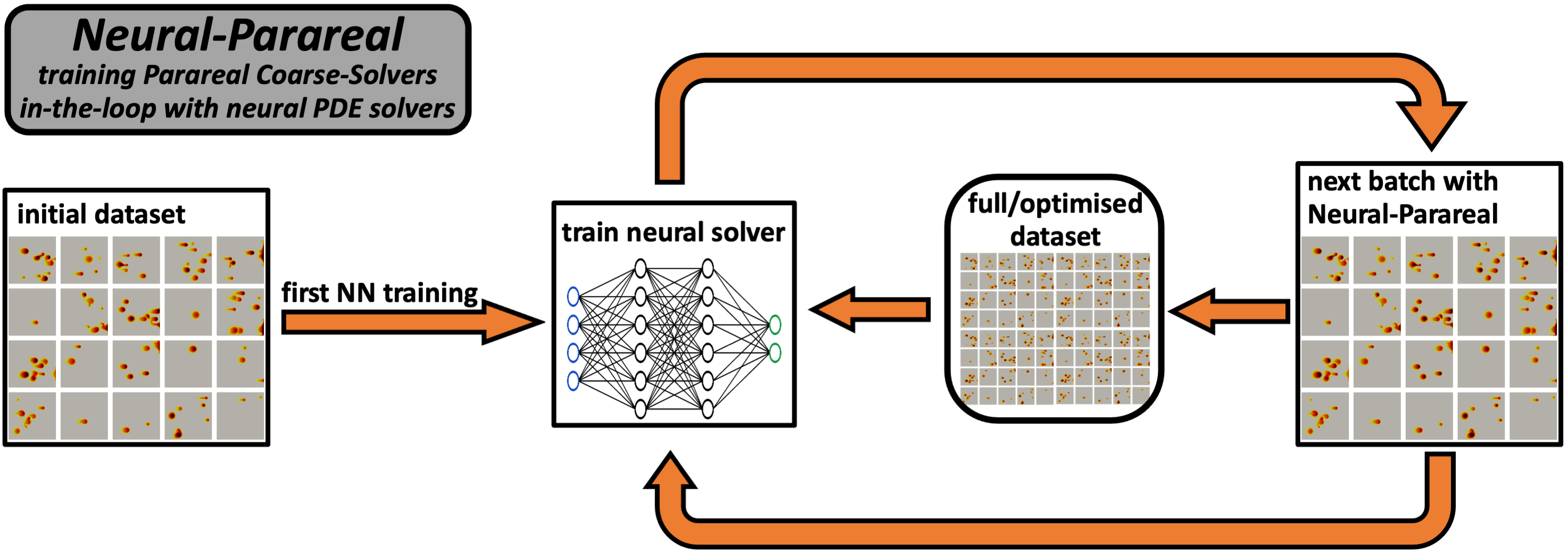}}
  \caption{\scriptsize\textit{\newline The Neural-Parareal framework, where neural operators are trained live as more simulation data becomes available, to provide a progressively more precise coarse-solver, thus leading to more potential speedup and accuracy as simulations are being produced.}}
  \label{FIGURE_Neural_Parareal}  
\end{figure}   

The closest work that the authors are aware of, and similar to the study presented here, is that of Gorynina et al. \cite{Gorynina2022}, where a Machine Learning surrogate is used as coarse solver for molecular dynamics, and more recently, of notable interest, the work by Qadir-Ibrahim et al. \cite{Qadir-Ibrahim2024}, where a Physics-Informed version of the FNO \cite{Z.Li-FNO-2020} algorithm (PINO \cite{Z.Li-PINO-2021}) is used as the coarse solver. Another relevant study is that of Pentland et al. \cite{Pentland2023_GParareal}, where a Gaussian Process is used to learn the difference (i.e. the predictor-corrector) between a given coarse-solver and the fine-solver. The way the work described here differs from previous achievements is both the application, which is fusion-specific with a set of highly non-linear PDEs (MHD), and most importantly the demonstration of how the training of the Machine Learning coarse solver can be integrated into the Parareal workflow to produce a more accurate coarse solver as the number of simulations increases. This framework integration can be compared to Solver-in-the-Loop methods \cite{Um2020,Toshev2024}, where the simulation code is integrated inside the neural operator training, except that it does the reverse: here the training of neural operators is integrated inside the HPC simulation algorithm. This `AI-in-the-Loop' idea is illustrated in Figure-\ref{FIGURE_Neural_Parareal}. It may be compared to methods where AI acceleration is used for preconditioning of simulations, such as \cite{Castagna2024}. This convergence of HPC and AI methods is most relevant to the development of Digital Twins and Digital Models for fusion research, where the need for rapid designs of future fusion power plants requires quick yet accurate estimations of costly and slow HPC simulations using Machine Learning surrogates.

\subsection{Fusion Application}

In the current alarming climate change situation \modifa{\cite{IPCC2021}}, nuclear fusion could provide an abundant energy source with a minimal level of greenhouse gas emissions and no long-lived radioactive nuclear waste. Together with renewable energies, fusion could contribute to the electricity of future societies, without the limit of exhaustible natural resources. Currently, the most promising candidate for industrial fusion reactors is the tokamak device \modifa{\cite{Wesson2004}}, which uses a magnetic field to confine a hot plasma of ionised hydrogen isotopes. The toroidal, periodic nature of the tokamak ensures that the hydrogen ions and the electrons, which approximately follow the magnetic field lines, are not lost at the end of open field lines, like in linear plasma devices. However, this periodicity can lead to resonance and instability. Resonant and unstable modes typically involve the plasma and the magnetic field, and are commonly studied using MHD models \cite{Freidberg2014,Schnack2006,Sauter1999}, combining the Navier–Stokes equations with Maxwell's equations.

Theoretical analysis of the MHD equations can provide some limited insight into the properties of various waves and unstable modes in a tokamak \cite{Freidberg2014}, however to obtain a more detailed understanding of tokamak MHD instabilities, numerical simulations are required. Some of the main tokamak MHD instabilities include Edge-Localised-Modes (ELMs), Toroidal Alfven Eigenmodes (TAEs) and Global instabilities (Disruptions). ELMs eject plasma filaments from the edge region onto the first wall of the machine, leading to large heat-fluxes on surface materials \cite{Eich2003,Sieglin2017,Ham2020,Leonard2014}. TAEs, which are excited by the 3.5MeV alpha-particles born of fusion reactions, can limit the performance of plasma operations \cite{Cheng1985,Dvornova2020,Fitzgerald2020,Pinches2015}. Global MHD instabilities which affect the entire plasma can lead to the total loss of plasma control, these are called disruptions. During disruption events, the kinetic and magnetic energy of the plasma can be transferred to the wall, leading to unsustainable material heat-fluxes and/or wall-currents that can damage the structural components of the machine \cite{Boozer2012,DeVries2016,Lehnen2015,Artola2020,Hu2018,Bandaru2019}. In order to study, understand and predict these MHD instabilities, numerical simulations are performed using codes like JOREK \cite{JOREK_overview,Huysmans2007,Czarny2008,JOREK}, M3D-C1 \cite{M3DC1,Jardin2004}, NIMROD \cite{NIMROD,Sovinec2004}, XTOR \cite{Lutjens2008}, BOUT++ \cite{BOUT++,Dudson2009}, MEGA \cite{Todo1998,Todo2012,Konies2018}, HALO \cite{Fitzgerald2020} (and many others).

In this study, we use the Reduced-MHD equations \cite{Strauss1997,Strauss1976} in a 2D slab geometry with toroidal curvature (toroidally axisymmetric domain). The simulations are evolving filamentary blobs similar to ELM filaments at the plasma edge. Although the geometry is simplified, the physics model and type of dynamics is similar to state-of-the-art applications of the JOREK code \cite{JOREK_overview,Pamela2020}, and therefore represents a practical demonstration from which future extensions of the framework could be developed to address realistic tokamak use cases. Note that blob convection in tokamaks is also extensively studied in the context of electrostatic turbulence, such as in \cite{Militello2017,Ross2019_Grillix_blobs}.

\subsection{Overview of the work}

In this paper, we present an integrated framework that combines Parareal simulations of the JOREK code \cite{JOREK} with the training of neural operators in PDEarena \cite{PDEarena,Jayesh2022}, bootstrapped into the workflow to benefit from the large data-generation of the Parareal algorithm in real time. This integration results in progressively more and more accurate coarse solvers as the input-parameter domain is explored with new simulations, and thus potentially higher speed-up. Section-\ref{SECTION_MHD} introduces the MHD simulation use cases that were used for the development of the Parareal framework and the initial development of the Neural Operator surrogates. Section-\ref{SECTION_NN} presents the work done with PDEarena \cite{PDEarena,Jayesh2022} to create surrogate models of the simulations, which are used as the coarse solvers of the Parareal framework. Section-\ref{SECTION_PARAREAL} describes the full implementation of the Parareal framework which accommodates the FEM discretisation of JOREK, the neural coarse solvers from PDEarena, and the non-negligible parallel i/o processing required for an efficient framework. Finally, Section-\ref{SECTION_RESULTS} presents the main results of the framework while Section-\ref{SECTION_CONCLUSION} summarises the work and lays out the further improvements desirable for future studies and extensions.


\section{The MHD Simulation Use Cases}\label{SECTION_MHD}

In this study, two use cases are addressed, both simulating blob convection in a 2D slab domain with toroidal axisymmetric geometry. The first use case employs a simplified electrostatic model, which already has a large dataset published in several studies \cite{Carey2024,Gopakumar2024}. This use case was used for the development of the Neural-Parareal framework and initial tests that the performance was reasonable.

The second use case is a new dataset based on similar blob simulations but with a more complex MHD model, the so-called Reduced-MHD model, which has been used extensively in literature for the study of tokamak instabilities \cite{JOREK_overview,Pamela2020}. This use case was used to demonstrate the integrated framework with the dynamic training of the coarse-solver, bootstrapped inside the workflow to exploit the data generated by new simulations.

In both use cases, the simulations are run with a bi-cubic (high-order) C1-continuous Bezier finite-element grid with uniform resolution of 200 by 200 elements. The poloidal 2D slab is centered at a toroidal major radius of $10m$ with height and width of $1m$. The time-step size is approximately $0.15\mu s$. Both spatial and temporal resolutions are voluntarily chosen to be conservatively high (fine) to ensure numerical stability across the entire input-parameter domain. For all simulations, 2000 time-steps are run. The boundary conditions around the domain are Dirichlet for all variables.

\subsection{Electrostatic Blob Simulations}\label{SECTION_MHD1}

The first use case employs an electrostatic model, which is equivalent to the Reduced-MHD model as routinely used in JOREK \cite{JOREK_overview}, but without the magnetic potential and current. There are a total of 4 variables in the model: 3 physical variables and an auxiliary variable, used for numerical stability (see \cite{JOREK_overview}). These physical variables are the fluid density $\rho$, the fluid temperature $T$ and the electric potential $\Phi$. The auxiliary variable is the toroidal vorticity, defined as $\omega = \nabla^2\Phi$. Note that the Laplacian here is in toroidal coordinates. This model is very similar to the Navier-Stokes equations, where $\Phi$ can be associated to the stream function of the fluid velocity. The exact formulation of the velocity is given, as in \cite{JOREK_overview}, with $\vec{v}=R^2\nabla\phi\times\nabla\Phi$, where $R$ is the major radius, and $\phi$ is the toroidal coordinate. One can easily derive that the toroidal vorticity is in fact simply $\omega=\nabla\phi\cdot(\nabla\times\vec{v})$.

The simulations are initialised with multiple blobs inside the slab, varying randomly the number of blobs, their positions, their (2D-Gaussian) width, their density amplitude and their temperature amplitude. The range of these input parameters are as follows:
\begin{flushleft}
\begin{tabular}{ |p{6cm}|c|p{4cm}|  }
 \hline
 \textbf{input parameter} & \textbf{min/max} & \textbf{type \& unit} \\ 
 \hline
 number of blobs & [1 : 10] & discrete \\ 
 R-position of blobs & [9.6 : 10.4] & continuous [$m$] \\ 
 Z-position of blobs & [-0.4 : 0.4] & continuous [$m$] \\ 
 width of blobs & [0.02 : 0.1] & continuous [$m$] \\ 
 density amplitude of blobs & [0.1 : 0.4] & continuous [$10^{20}m^{-3}$] \\ 
 temperature amplitude of blobs & [12 : 72] & continuous [$eV$] \\ 
 \hline
\end{tabular}
\end{flushleft}
These quantities are representative of small filamentary blobs in the Scrape-Off Layer of a tokamak plasma, i.e. just outside the hot, confined plasma region. In a tokamak, such filamentary structures are expelled from the confined region due to turbulence and/or MHD instabilities. As a reference, the electron density and temperature in the JET-ILW tokamak, just at the edge of the confined plasma region, is typically of the order of $5.10^{19}m^{-3}$ and $1keV$.

The electric potential $\Phi$ is initialised as zero. As the simulation starts, the toroidal curvature combined with the pressure gradient of the blobs generates an electric field that leads the blobs to move radially outwards (away from the centre of the torus). The hotter the blob, the faster its motion. The Dirichlet boundary conditions cause the blob material to mix inside the slab until they dissipate through diffusion.

\begin{figure}[ht!]
  \centering
	\subfigure[ ]{\includegraphics[width=16.0cm]{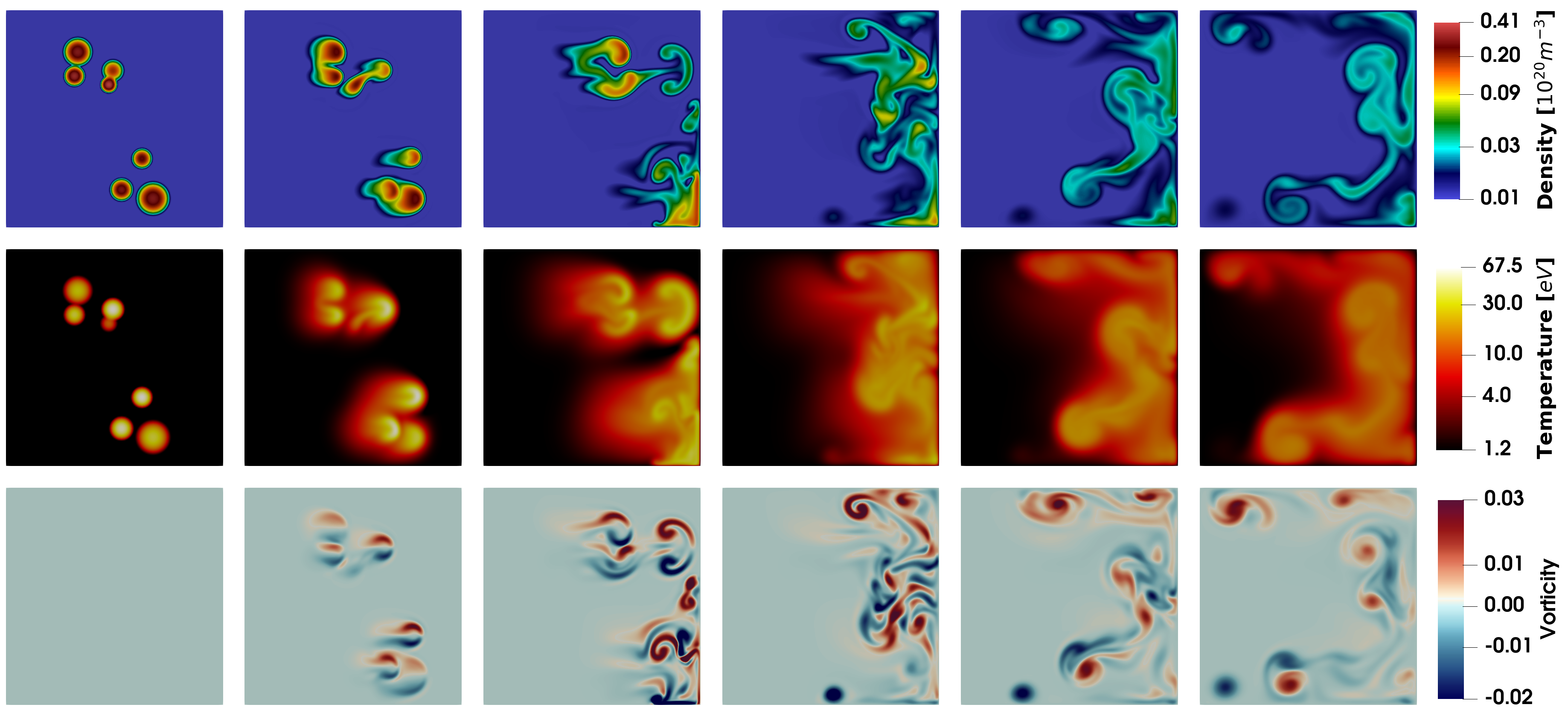}}
  \caption{\scriptsize\textit{\newline The first use case with an electrostatic Reduced-MHD model, showing one of the simulations with blobs initialised in a poloidal 2D slab domain. The first 1000 time-steps are illustrated here (half of the full simulation), with each frame corresponding to approximately $30\mu s$, i.e. [0,30,60,90,120,150]$\mu$s.}}
  \label{FIGURE_BLOB_3VAR}  
\end{figure}   

The poloidal diffusion parameters are the density diffusion $D=3.5m^2.s^{-1}$, the temperature diffusion $\kappa=2.10^{-7}kg.m^{-1}.s^{-1}$, and the viscosity $\mu=2.10^{-6}kg.m^{-1}.s^{-1}$. For more details on the Reduced-MHD model and its parameters, see \cite{JOREK_overview}. Figure-\ref{FIGURE_BLOB_3VAR} shows an example of a simulation with 7 blobs moving radially towards the outer boundary of the domain. Note that only half the simulation (1000 time-steps) is shown in Figure-\ref{FIGURE_BLOB_3VAR} for illustration purposes. All datasets can be downloaded from Zenodo, including GIFs \cite{Zenodo_dataset_RMHD,Zenodo_dataset_3var_batch-0001-0500,Zenodo_dataset_3var_batch-0501-1000,Zenodo_dataset_3var_batch-1001-1500,Zenodo_dataset_3var_batch-1501-2000}. The simulations (2000 time-steps) take just above 6 hours to run on a 2x24-cores Intel-Xeon-8160 (SkyLake) node. More information about the existing (spatial) parallelisation of the JOREK code can be found in \cite{JOREK_overview}.

\subsection{Electromagnetic Blob Simulations}\label{SECTION_MHD2}

The second use case is exactly the same as above but with the more complex Reduced-MHD model, as implemented in routine JOREK studies, which includes a magnetic field but without the (optional) parallel velocity \cite{JOREK_overview}. This model has the same variables $\rho$, $T$, $\Phi$ and $\omega$ as before, with an extra physical variable for the poloidal magnetic potential $\psi$, and an additional auxiliary variable for the toroidal plasma current density, defined as $j=R^2\nabla(R^{-2}\nabla\psi)$. The magnetic field is given, as in \cite{JOREK_overview}, by $\vec{B}=\vec{B}_{\phi} + \vec{B}_{pol}$, where the toroidal magnetic field $\vec{B}_{\phi}=F_0\nabla\phi$ is constant in time in the Reduced-MHD model, and only the poloidal magnetic field $\vec{B}_{pol}=\nabla\psi\times\nabla\phi$ is evolved through the scalar potential variable $\psi$. Note that this is why the model is called \textit{Reduced}-MHD, as opposed to \textit{full}-MHD where the toroidal field also evolves \cite{Pamela2020}. In this set-up, one can easily derive that the toroidal current is in fact simply $j=R^2\nabla\phi\cdot(\nabla\times\vec{B})$.

\begin{figure}[ht!]
  \centering
	\subfigure[ ]{\includegraphics[width=16.0cm]{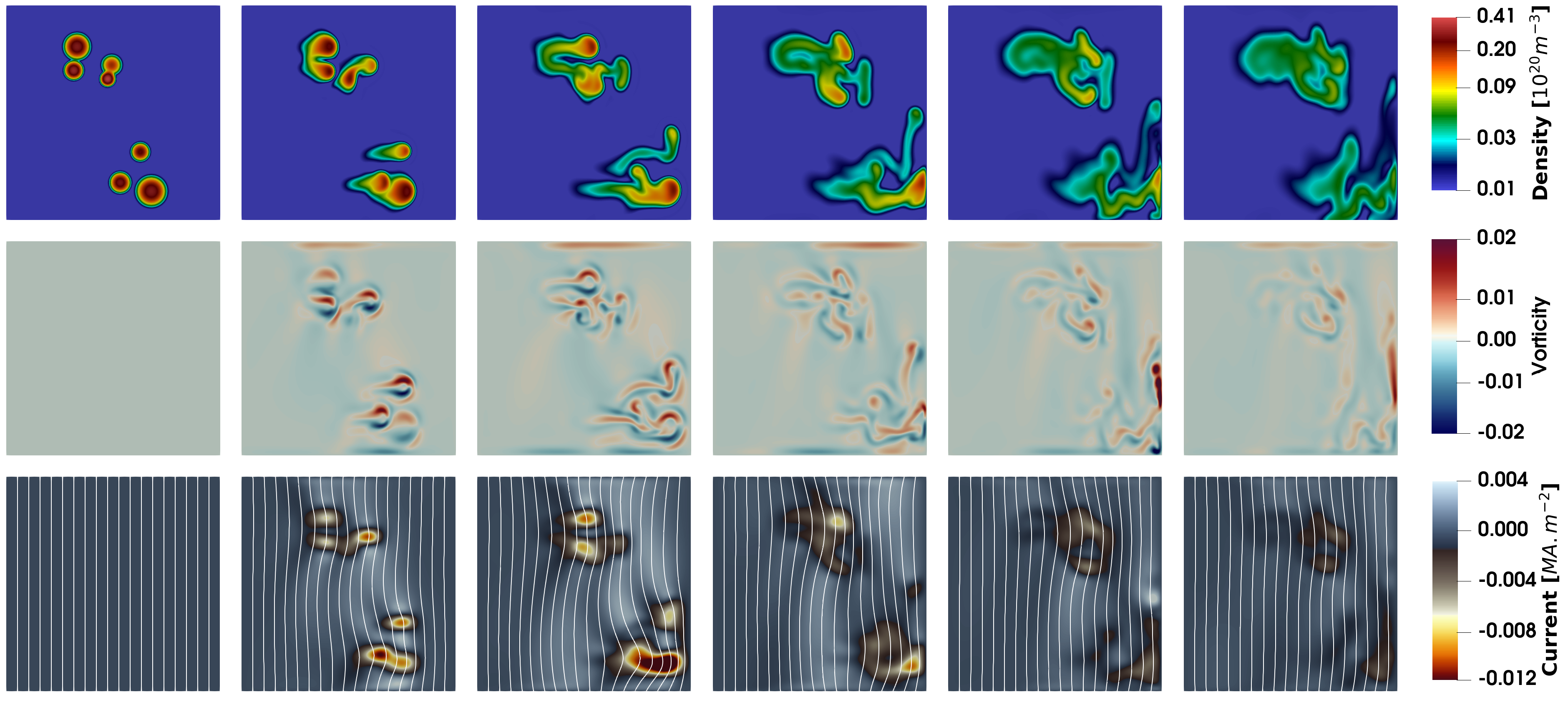}}
  \caption{\scriptsize\textit{\newline The second use case with an electromagnetic Reduced-MHD model, showing a simulation with the same initialisation as in Figure-\ref{FIGURE_BLOB_3VAR}. The first 1000 time-steps are illustrated here as well, each frame corresponding to approximately $30\mu s$. It is worth noting the difference between the electromagnetic model and the electrostatic model: the blobs move slower and deviate up/down-wards, due to the interaction with the background magnetic field. In the bottom row, the color shows the current generated by the filaments, as well as contours of the magnetic potential $\psi$, to illustrate the bending of the magnetic field by the blobs.}}
  \label{FIGURE_BLOB_5VAR}  
\end{figure}   

In the simulations presented here, the toroidal magnetic field is set to $1T$, and the poloidal magnetic field is set to a background of $10^{-3}T$ to represent the Scrape-Off Layer of a tokamak, just outside the confined plasma, where there the poloidal magnetic field is much lower than inside the confined plasma region. The poloidal magnetic field actually vanishes in the so-called X-point (saddle point) region of the plasma.

No current is initialised inside the blobs, but as can be seen in Figure-\ref{FIGURE_BLOB_5VAR}, as the blobs start evolving, they generate their own current, which affects their dynamics and bends the magnetic field. This is illustrated by the bottom row of the figure, which shows the current scalar together with contour lines of $\psi$ (which can be considered as the stream function of the poloidal magnetic field). The resistivity in these simulations is relatively high, at $4.10^{-5}\Omega.m$, such that the plasma fluid is not `frozen' to the magnetic field and can travel through flux surfaces. Still, the field-line bending and current have a significant effect on the dynamics of the blobs, which can be clearly seen when comparing Figure-\ref{FIGURE_BLOB_5VAR} to Figure-\ref{FIGURE_BLOB_3VAR} which has exactly the same initial conditions. In particular, with the Reduced-MHD model, the blobs can be observed to travel slower and deviate up/down-wards compared to the simpler model where they just travel radially outward until reaching the outer boundary.

With this Reduced-MHD model, the simulations (2000 time-steps) take just under 14 hours to run on a 2x24-cores Intel-Xeon-8160 (SkyLake) node. Note that in a fully implicit time-scheme, the problem (matrix) size scales as the number of variables squared, so the factor 2 in computation increase between the two models corresponds to $6^2/4^2=2.25$.


\section{Neural Operator Surrogates}\label{SECTION_NN}

\subsection{PDEarena using FNO method}

The initial dataset using the electrostatic model from Section-\ref{SECTION_MHD1} was used for previously published studies in \cite{Gopakumar2024,Carey2024}. It comprises of 2000 simulations \cite{Zenodo_dataset_3var_batch-0001-0500,Zenodo_dataset_3var_batch-0501-1000,Zenodo_dataset_3var_batch-1001-1500,Zenodo_dataset_3var_batch-1501-2000}. From these samples, 90\% was used for training, and 10\% for testing. Because conservative resolution was used in the simulations, the data is down-sampled both spatially and temporally before ingestion into the neural operator training. The spatial resolution is set to 100$\times$100, while the temporal frequency is reduced by a factor 10, hence 200 time-frames per simulation. All variables are normalised to [-1:1] with respect to the minima/maxima of the entire dataset.

A neural operator learning \cite{Lu:19, Lu:21, Li:20graph, Li:20,Kovachki:21} framework is constructed to learn a mapping between function spaces – as needed when approximating solutions of partial differential equations (PDEs). Similar to \cite{Kovachki:21}, it assumes $\mathcal{U}, \mathcal{V}$ to be Banach spaces of functions on compact domains $\mathscr{X} \subset \mathbb{R}^{d_x}$ or $\mathscr{Y} \subset \mathbb{R}^{d_y}$, mapping into $\mathbb{R}^{d_u}$ or $\mathbb{R}^{d_v}$, respectively. The goal of operator learning is to learn a ground truth operator $\mathcal{G} : \mathcal{U} \rightarrow \mathcal{V}$ via an approximation $\hat{\mathcal{G}}: \mathcal{U} \rightarrow \mathcal{V}$. This is usually done in the vein of supervised learning by independent and identically distributed (i.i.d.) sampling input-output pairs, with the notable difference that in operator learning the spaces sampled from are not finite dimensional. More precisely, with a given data set consisting of $N$ function pairs $(\mathbf{u}_i, \mathbf{v}_i)=(\mathbf{u}_i, \mathcal{G}(\mathbf{u}_i) )\subset \mathcal{U} \times \mathcal{V}$, $i=1,...N$, a neural operator aim to learn  $\hat{\mathcal{G}}: \mathcal{U} \rightarrow \mathcal{V}$, so that $\mathcal{G}$ can be approximated in a suitably chosen norm.

The PDEarena platform \cite{PDEarena,Jayesh2022} was used to train the surrogate models. Although PDEarena includes several options of neural operators, in this work only the Fourier-Neural Operator method (FNO) \cite{Z.Li-FNO-2020} was used. PDEarena was used as a code base and extended to support the JOREK simulation data. A modified version of the FNO configuration `FNO-128-32m' in PDEarena is used with the number of Fourier blocks increased to 3, and where the grid discretisation is concatenated in the same dimension as the physics variable fields as it was found to improve performance, as demonstrated in \cite{Gopakumar2024}.

The FNO method \cite{Z.Li-FNO-2020} trains a neural network in both the real space and a Fourier space representation of the data (for a given number of Fourier modes). This effectively means the neural network learns a functional mapping of the training data rather than just discrete data values. The advantage of this method is that any interpolation between data points (both spatially and in terms of the input domain) with be more reliable than for data-only neural networks. The FNO has been shown to work well on a wide number of applications, including fluid models with convection particularly relevant to the use case presented here.

\subsection{Rollout and network inputs}

For a given temporal resolution, the neural operator is trained to predict $k$ time-steps ahead, given $l$ time-steps as input. These parameters $k$ and $l$ will affect the performance of the predictions, but they also affect how the solver can be integrated into the Parareal framework, which will be addressed in more details in Section-\ref{SECTION_PARAREAL}. Typically, and in all cases included in this study, $l$ is chosen to be $1$, while $k$ is varied between 5 and 20. 

In order to predict far ahead of a given set of input time-steps, the prediction is \textit{rolled out} in an autoregressive manner. Given [1$\xrightarrow{}$$k$] input steps, once the step $k$+1 has been predicted, a new set of inputs [2$\xrightarrow{}$$k$+1] is fed to the network to predict the step $k$+2. Following which the inputs [3$\xrightarrow{}$$k$+2] are fed back into the network to predict step $k$+3, and so on until the desired prediction length is achieved. This is illustrated in Figure-1 of \cite{Carey2024}.

If $\mathcal{G}$ is the mapping from an initial condition $\mathbf{u}(0,\mathbf{x})=\mathbf{u}^0(\mathbf{x})$ to the solutions $\mathbf{u}(t,\mathbf{x})=\mathbf{u}^t(\mathbf{x})$ at later times, then in order to obtain accurate predictions over long time horizons, a temporal operator could either be directly trained for large $\Delta t$ or recursively applied for smaller time intervals. However, in practical applications, the predictions of neural operators degrade for large $\Delta t$, while autoregressive approaches are found to perform substantially better \cite{Gupta:22, Li:22, Wang:23, Lippe:23}.

The models are trained for about 72 hours (depending on performance) using the Adam optimizer with cosine annealing learning rate scheduler with the starting learning rate of 0.0002 and minimum learning rate value of 1.e-7 for both. The number of epochs and learning rate was varied for optimum performance in some of the training runs, but systematic hyperparameter tuning of the entire model was not performed here and is being considered for future work, as described below.

\begin{figure}[ht!]
  \centering
	\subfigure[ ]{\includegraphics[width=16.0cm]{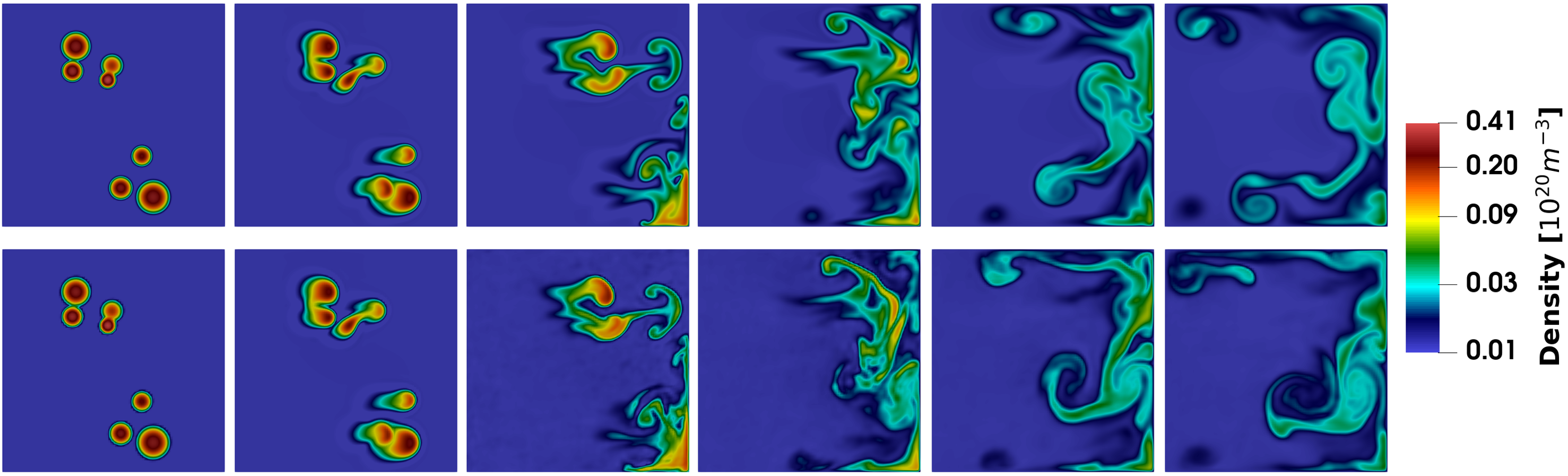}}
  \caption{\scriptsize\textit{\newline Comparison of the evolution of the real simulation (top) compared to the PDEarena surrogate (bottom). As in Figure-\ref{FIGURE_BLOB_3VAR}, each frame corresponds to 200 time-steps, so 30$\mu$s.}}
  \label{FIGURE_PDEarena_prediction}  
\end{figure}   

Again for the same simulation example as in Figure-\ref{FIGURE_BLOB_3VAR} (for the electrostatic model), the prediction of the PDEarena surrogate is rolled out and compared to the ground truth in Figure-\ref{FIGURE_PDEarena_prediction}. Each frame of Figure-\ref{FIGURE_PDEarena_prediction} corresponds to 200 time-steps (30$\mu$s) from the original simulation, thus 20 steps of the PDEarena temporal resolution. For this example, the neural solver was trained using 20 steps as input, meaning that the second frame is, in fact, still the actual simulation. The next 4 frames, 80 steps in total, are thus entirely predicted by the neural solver. As can be seen, the precision of the prediction is reasonably reliable for about 40 steps, beyond which it diverges from the ground truth.

The run-time of neural operators like the FNO is one of the main points of this study: it is extremely cheap. For a full rollout of the same length as a simulation of 2000 time-steps (i.e. 200 rollout steps), the execution time is approximately 2 minutes on a single core. Most of the execution time is in fact dominated by the NN-model load. At present the model is loaded each time an evaluation is needed, but future improvement may include having background jobs with the model loaded and awaiting external signals to process model evaluations whenever needed. This may sound excessive, considering that 2 minutes is negligible in comparison to the real simulations that require 14 hours on 48 cores, but in a Parareal framework, the speed of the coarse-solver is essential. For example, if a Parareal run is executed with 100 time-windows, and each coarse-solver evaluation takes 2 minutes, it quickly adds up.

\subsection{Limitations of current versions}

More work is currently under way to improve the PDEarena solvers, both for the purpose of exploration of surrogates in fusion applications in general, but also for this specific application with Parareal. Although these improvement areas are beyond the scope of this study, they are worth mentioning here for the sake of clarity.

Firstly, using higher spatial and temporal resolution may have significant effects on the precision of the neural solver. Although 200$\times$200 bi-cubic finite-elements may be conservative, the granularity of the 100$\times$100 grid used for the neural solver can be seen by eye when zooming on the details of Figure-\ref{FIGURE_PDEarena_prediction}. With a set of non-linear PDEs, any loss of precision will undoubtedly accumulate to significant deviation for long rollout predictions. Likewise, down-sampling the temporal frequency of the data may play a role. Of course, increasing resolution means the training time and cost of the models would increase significantly, which is part of the reason why this is being kept for future plans. At present, parallelising the training on multiple GPU nodes is being explored to alleviate this limitation.

Secondly, there are more neural operator options in PDEarena, besides the FNO, which ought to be explored. The efficiency of each of these methods may be strongly dependent on the use case. For example, one particular model may perform very well on 2D regular meshes, but could become unreliable when addressing unstructured meshes in 3D with more complex geometries. The most relevant aspect of this issue is to scale up towards realistic fusion applications, such as turbulence in 3D toroidal geometries, which will require larger models and larger datasets, for which transformer architectures may be most appropriate.

Finally, hyperparameter tuning has not been done systematically in this study, as it represent another dimension to the total cost of the framework as a whole. Note that it isn't just the internal parameters of each model that should be subject to optimisation, but the models themselves, as well as the data-resolution described above. It's entirely possible that, depending on the use case, as more data is made available to the neural solver training, different models and different resolutions may be more appropriate, not just model parameters.


\section{Parareal Framework}\label{SECTION_PARAREAL}

Although the implementation of the Parareal algorithm is generally straight-forward, in this particular case there are two aspects that make the framework more complex than a standard situation. In particular, the fact that the JOREK code uses finite-elements, and that the neural coarse solver to be used requires several input time-steps, as opposed to a single initial-value state in typical Parareal applications. 

\subsection{Parareal with a finite-element fine-solver}

As a first demonstration, the Parareal framework was first implemented using a \textit{classical} coarse-solver, namely the same simulation code JOREK, but with (optionally) coarser spatial/temporal resolution. This step was useful not just to develop the framework itself, but as will be seen in Section-\ref{SECTION_RESULTS}, it also gives a practical reference for testing and evaluation.

The first technical aspect in the implementation is to convert data from one spatial resolution to another. In this context, given two equidistant grids of point-wise data with different resolutions, simple algorithms like linear interpolation are easily applied. However, when finite-elements are involved, such conversions need to be \textit{projected} onto the degrees-of-freedoms of each element. This projection, which is applied for each scalar variable independently, requires the weak-form integration of each element around a given node to solve for its degrees of freedom. This procedure is already an available feature in the JOREK code, but it requires the data to be evaluated at the Gaussian integration points of each element. For bi-cubic elements, 4 Gaussian integration points in each direction are necessary for each element. Thus, whether data is being up-sampled or down-sampled between two grids, if the final destination is a finite-element grid, this sampling must be done on the Gaussian integration points. It is important to note that the location of Gaussian integration points are not equidistant, such that linear interpolation is not adequate.

Whether Parareal is run with a \textit{classical} coarse-solver that uses finite-elements or not, interpolation between different resolutions is clearly required, and since these interpolations involve non-equidistant Gaussian integration meshes (and with the long-term goal of extending the framework to unstructured grids), a robust and generic approach is to use the Clough-Tocher 2D-Interpolator from the \texttt{scipy.interpolate} library. This method has the great advantage that it is mesh-agnostic, thus ideal for this application.  

Consider a fine-solver $F$ and coarse solver $G$ that evolve on different grid resolutions. In practice, it is safe to assume that the fine-solver has the highest spatial resolution. Note that although this is not strictly necessary, one would assume that if the coarse-solver also has high spatial resolution, it is because its grid is the same as the fine-solver, in which case no interpolation is needed. At each Parareal cycle $i_p$ and time-window $i_t$, the predictor-corrector algorithm must be applied to the outputs from the previous time-window and previous Parareal cycle, such that the new initial-value map is given by
\be
  U|_{(i_p,i_t)} = G|_{(i_p,i_t-1)} + F|_{(i_p-1,i_t-1)} - G|_{(i_p-1,i_t-1)}. \label{EQ_predictor_corrector}
\ee
However, it is important to note that this operation involves data from two different grids as inputs, and that the output $U|_{(i_p,i_t)}$ must be evaluated on both the coarse grid and the fine grids in order to run the next Parareal cycle. Thus there are two choices:
\begin{enumerate}[label={(\alph*)}]
    \item all inputs $G|_{(i_p,i_t-1)}$, $F|_{(i_p-1,i_t-1)}$ and $G|_{(i_p-1,i_t-1)}$ are first interpolated onto the higher-resolution grid, the predictor-corrector operation (\ref{EQ_predictor_corrector}) is applied, and the resulting initial-value map $U|_{(i_p,i_t)}$ is interpolated onto the coarse grid. \label{IO_option-a}
    \item all inputs $G|_{(i_p,i_t-1)}$, $F|_{(i_p-1,i_t-1)}$ and $G|_{(i_p-1,i_t-1)}$ are interpolated onto both the fine-solver grid and the coarse-solver grid, and the predictor-corrector operation (\ref{EQ_predictor_corrector}) is applied to both sets of interpolated inputs. \label{IO_option-b}
\end{enumerate}

While the difference between these two option may seem irrelevant, there is one important point to consider: at each Parareal cycle $i_p$, and at each time-window $i_t$, the predictor-corrector step can only be applied once the coarse-solver solution has been obtained for the previous time-window $i_t$$-$$1$. In other words, the predictor-corrector step is sequential across time-windows, just like the coarse-solver evaluation. However, interpolations with the Clough-Tocher method can be non-negligible, particularly when running realistic use cases with high resolution grids. Note, it is most expensive to interpolate \textit{from} a high-resolution grid, but once the Clough-Tocher spline has been calculated, evaluation is relatively cheap. This means that for option-\ref{IO_option-a}, the final interpolation onto the coarse-solver grid will be expensive. The advantage of the second option-\ref{IO_option-b} above is that the interpolation of all $F$ and $G$ solutions, on both coarse- and fine-solver grids, can be executed at the end of each evaluation independently. That means the most expensive interpolation (\textit{from} $F$) can be done in parallel when the fine-solver evaluation is deployed. And only the (cheaper) interpolation of $G|_{(i_p,i_t-1)}$ remains sequential since the interpolation of $G|_{(i_p-1,i_t-1)}$ is already available from the previous cycle. 

\begin{figure}[ht!]
  \centering
	\subfigure[ ]{\includegraphics[width=16.0cm]{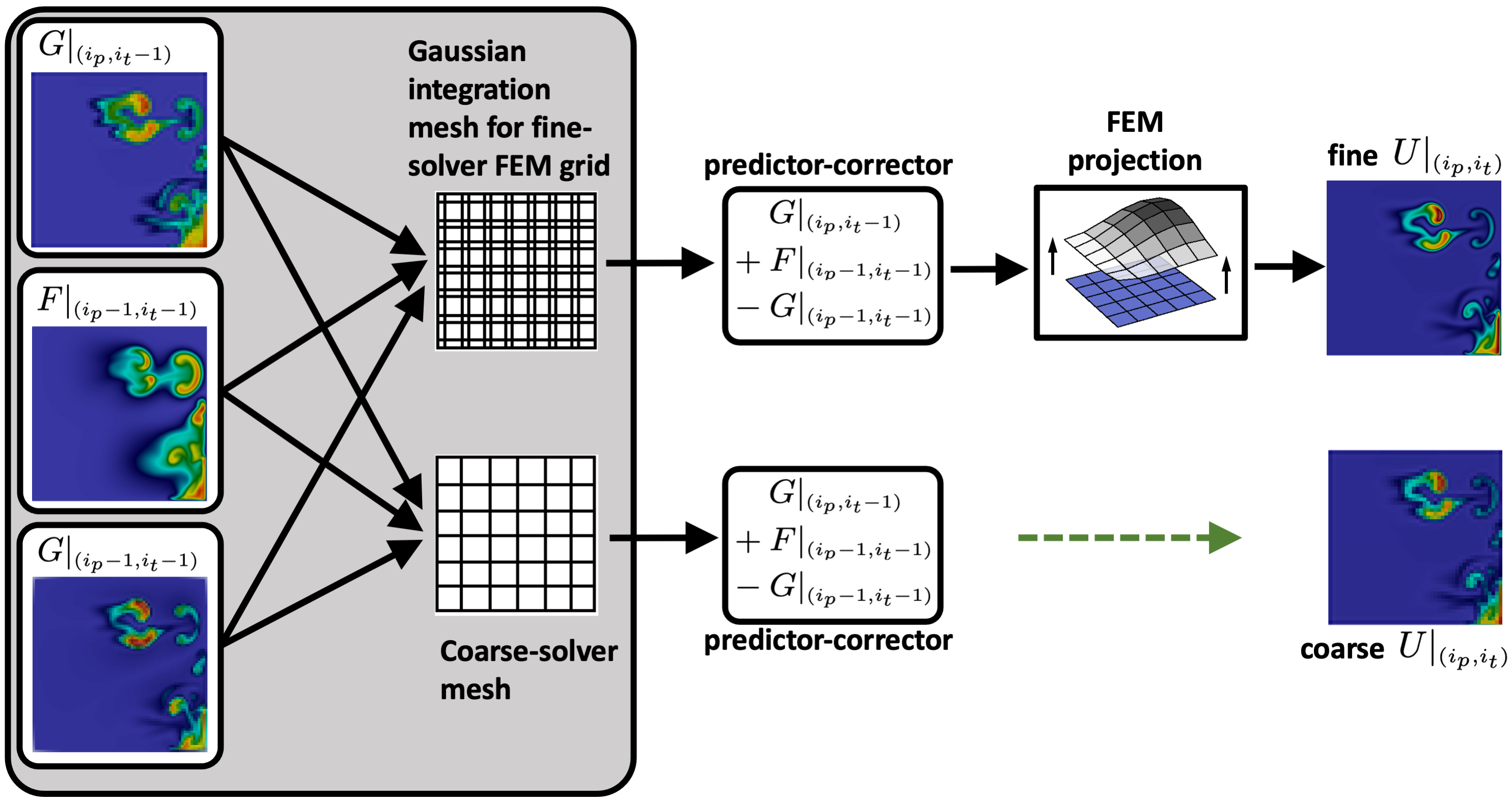}}
  \caption{\scriptsize\textit{\newline Schema of the Parareal predictor-corrector step for an application with finite-element grids. Note that if the coarse-solver also uses a finite-element grid (such as the `classical' coarse-solver), the last step (dashed-green arrow) also involves a FEM projection. In order to avoid i/o overheads at run-time, the evaluation/interpolation onto the two grid domains (grey box) is executed together with the deployment of each $F|_{(i_p-1,i_t-1)}$ and $G|_{(i_p-1,i_t-1)}$. This is particularly important for the (most expensive) interpolation of the fine-solver $F|_{(i_p-1,i_t-1)}$, which can therefore be run in parallel, rather than sequentially at the end of each time-window's coarse-solver pass $G|_{(i_p,i_t-1)}$. Note that the 2D frames of the blobs in this figure are representative, and the coarseness is exaggerated on purpose to illustrate the coarse-solver}}
  \label{FIGURE_predictor-corrector}  
\end{figure}   

There is one additional aspect to this interpolation issue. The interpolation \textit{from} a given fine-solver input $F|_{(i_p-1,i_t-1)}$ does not necessarily have to use the Clough-Tocher method (or any interpolation method), since the fine-solver grid is in fact already a bi-cubic spline itself. This means that the $F|_{(i_p-1,i_t-1)}$ inputs can be directly \textit{evaluated} onto the required grids (either coarse-solver or Gaussian-integration points), and no spline calculation is required. Note, however, that this step is fast provided the local-coordinates $(s,t)$, in finite-element space, are already known for each evaluation point. This means that for each evaluation point with poloidal coordinates $(R,Z)$ in real-space, the corresponding FEM local-coordinates $(s,t)$ have to be calculated and recorded at the beginning of each Parareal run, since these evaluations will be repeated for each time-window at each Parareal cycle (i.e. potentially thousands of times). This local-coordinates mapping is achieved using the Newton-Raphson method for each point and the resulting map saved for future evaluations. Contrary to this optimal approach, if option-\ref{IO_option-a} is employed, then the final interpolation of the high-resolution $U|_{(i_p,i_t)}$ has to be done using an interpolator like the Clough-Tocher method, since that $U|_{(i_p,i_t)}$ map is already a mixture of 3 solutions on a grid that isn't a spline (typically the Gaussian-integration mesh of the high-resolution finite-element grid).

Option-\ref{IO_option-a} is manageable for small tests and code-development examples but for realistic applications it becomes prohibitively expensive, such that the interpolation i/o can even dominate over everything else. This becomes particularly problematic for very large numbers of time-windows, where the evaluation of each fine-solver becomes faster (fewer time-steps to compute) but where the number of predictor-corrector operations increase.

Nevertheless, as will become evident in the next section, there is one disadvantage to option-\ref{IO_option-b}. Namely, the more natural implementation of option-\ref{IO_option-a} is doing all the data i/o and processing `on-the-fly', meaning that for each predictor-corrector operation, all the mappings required (on fine- and/or coarse-grids) can be generated instantly, even in-memory, and discarded as soon as the output of the predictor-corrector is obtained. With option-\ref{IO_option-b}, that data must be saved on the file-system and saved until at least the end of the next Parareal cycle, since the predictor-corrector operation requires the interpolated data from $F|_{(i_p-1,i_t-1)}$ and $G|_{(i_p-1,i_t-1)}$.

\subsection{Parareal with a neural coarse solver}

In the case where a neural coarse solver is used, like the FNO method in PDEarena, the input data needed for each coarse-solver run, at each time-window, is not just a single initial-value state. As described in Section-\ref{SECTION_NN}, the neural solver requires several time-steps as input. In practice, this means that the predictor-corrector step described above must be applied for several time-steps.

While this may sound like a simple matter of iterating the predictor-corrector step over each input-step, in practice this characteristic of Neural-Parareal has several implications, both in terms of data-processing and data-management, as explained below.

\subsubsection{Checkpoint synchronisation}

Firstly, this multiple-input constraint implies that the correct frequency of checkpoints is required for both the fine-solver and coarse-solver, to coincide with the required data-input of the neural solver, compared to standard Parareal applications, where only the total duration of the time-window needs to be the same for both solvers, with only one final checkpoint for each time-window. Note that checkpoint here refers to a \textit{state} file, or a \textit{restart} file which holds all the information necessary from which to continue the simulation.

Additionally, since the neural coarse-solver requires multiple input checkpoints, the Neural-Parareal framework cannot be run like a standard simulation with an initial condition. It requires a first \textit{pre-run} time-window with the fine-solver to create the initial set of multiple input checkpoints, from which the full Parareal simulation is initiated.

\subsubsection{Performance and i/o parallelisation}

As mentioned above, each predictor-corrector operation is not negligible, and if it needs to be run for several checkpoints, possibly 10 to 20 of them, then this operation must be parallelised. Note that in the predictor-corrector schema described in Figure-\ref{FIGURE_predictor-corrector}, there are two locations where parallelisation is required. 

The interpolation/evaluation of each input, as described in the grey box of Figure-\ref{FIGURE_predictor-corrector}, can be run in parallel at the end of each coarse-solver and fine-solver run, simply deploying across however many time-steps are required by the neural solver.

The predictor-corrector step, however, occurs sequentially after each coarse-solver pass through the time-windows of each new Parareal cycle. For the fine-solver (top row of Figure-\ref{FIGURE_predictor-corrector}, only a single input is required (the junction between time-windows), but for the coarse-solver (bottom row), the predictor-corrector must be applied to all time-steps required as input for the neural solver. This operation must also be parallelised.

\subsubsection{Data generation (and annihilation)}

Since all operations to generate the mappings of the data onto the Gaussian-integration points of the fine-solver grid and the coarse-solver grid (grey box in Figure-\ref{FIGURE_predictor-corrector}) are not happening `on-the-fly' as the predictor-corrector and must be saved onto the file-system, that data can be deleted. 

Apart from the data generation/annihilation necessary for the predictor-corrector, the particularity of the Parareal algorithm is that it creates a lot more data than a normal simulation. Assuming that the framework is run with a large number of time-windows, then each Parareal cycle includes the equivalent amount of fine-solver time-steps as the entire run would (except that they are disjointed since they start with separate initial-conditions). Therefore, even if the Parareal run achieves the desired level of precision/convergence after 4 cycles, it means that approximately 4 times more data has been produced than for a single simulation. This large amount of extra data per run is ideal for training the neural operator.

A normal JOREK simulation (without Parareal) with 200 checkpoint files out of 2000 timesteps is about 5.8GB. This contains the full spatial resolution with all variables on the high-order finite-element grid. A full simulation (without Parareal) in the down-sampled format for the neural operator training (100$\times$100 grid, 200 time frames) is 93MB. A Parareal simulation with 20 time-windows, run for the full 20 Parareal iterations, and saving all of the intermediary data (grey box in Figure-\ref{FIGURE_predictor-corrector}), is about 265GB. After clean-up, the same Parareal simulation, without saving all these intermediary files, and keeping only strategic data, is about 61GB.

\subsubsection{Implementation with Slurm scheduler}

All runs were executed on HPC clusters that operate with a Slurm Workload Manager. The current implementation requires a Slurm job `master' that orchestrates all the Parareal run, its i/o and the deployment of children Slurm jobs. The master job submits a new Slurm job for each fine-solver time-window, but the coarse-solver time-windows can either be submitted as separate Slurm job or be executed inside the master job, assuming it does not require too many resources. This can be advantageous in case queuing times are elevated, since the master job will remain idle while waiting for its children jobs to return.


\section{Neural-Parareal Results}\label{SECTION_RESULTS}

\subsection{Initial demonstration and tests}

In order to validate the implementation of the Parareal framework, a series of tests were run. The first result is a test that the framework works and runs to completion. By definition, the Parareal algorithm can be run for as many cycles as there are time-windows. At each new cycle, the first time-window becomes the actual fine-solver target simulation. At cycle 1, time-window 1 will be run with the fine-solver starting from the initial conditions. At cycle 2, time-window 1 does not need to be run at all anymore, and time-window 2 will be run with the fine-solver, starting from where time-window 1 finished, i.e. the exact simulation is continuing. At cycle 3, time-window 3 becomes the continuation of the real simulation, and so on. Thus, at cycle $n$, where $n$ is the number of time-windows, the entire simulation has been run from start to end. In practice, one would never run all cycles of a Parareal simulation, since that becomes at least as slow as the simulation itself (i/o and coarse-solver timing added), but it is computationally much more expensive, namely $\sum_{i=1}^{n}\frac{i}{n}$ = $\frac{1}{2}(n+1)$ times more expensive, since at each cycle $i$, there are $n-i$ time-windows to be run, each costing $\frac{1}{n}$ times the cost of the full simulation. Nevertheless, for testing and demonstration, it is useful to be able to compare the Parareal to the real simulation. For this purpose, the last time-step of the last time-window of the Parareal simulation should converge towards the last time-step of the full simulation. If that convergence occurs quickly, it means the Parareal simulation is efficient.

\begin{figure}[ht!]
  \centering
	\subfigure[ ]{\includegraphics[width=16.0cm]{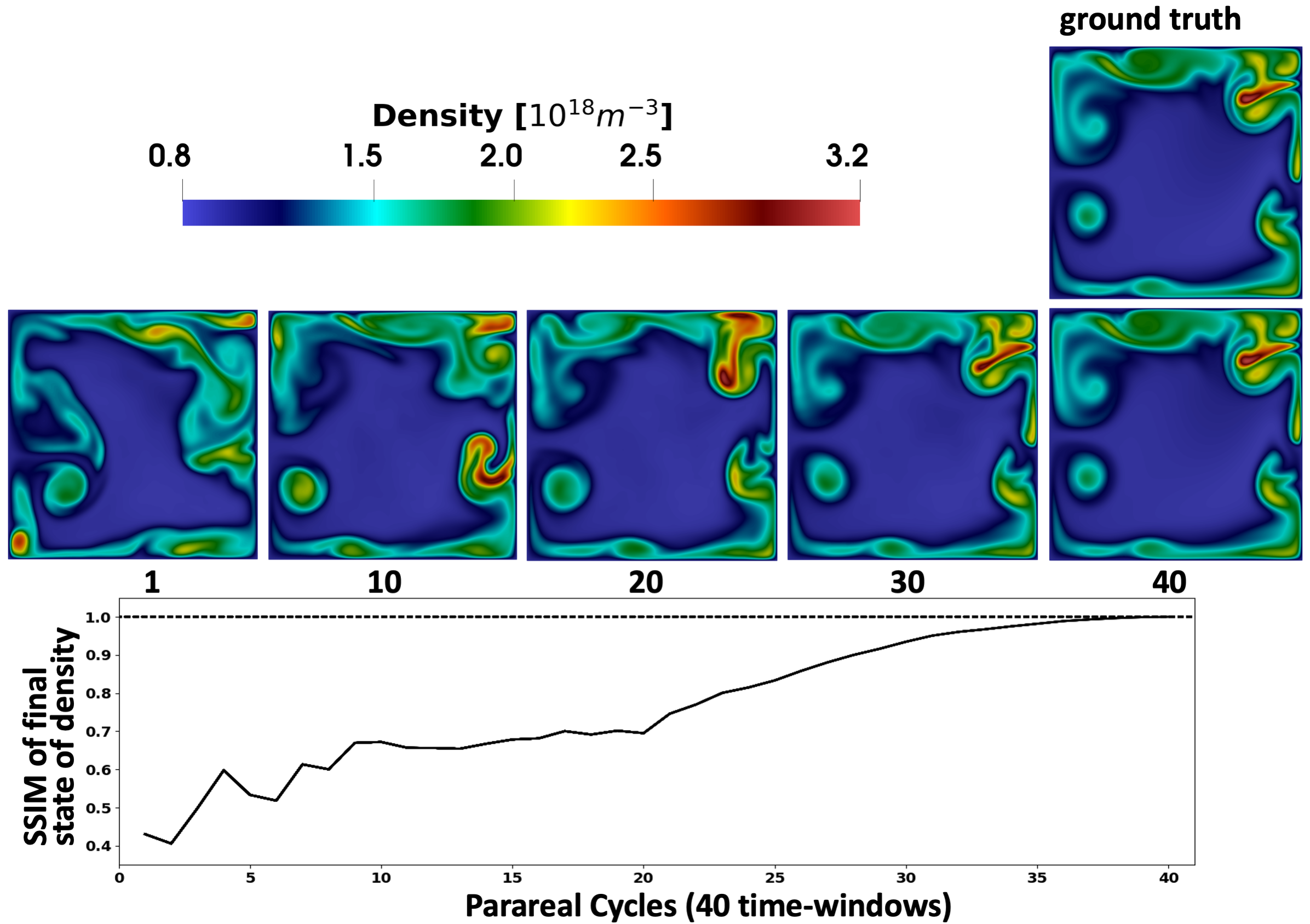}}
  \caption{\scriptsize\textit{\newline Evolution of the last time-step of the last (40$^{th}$) time-window of a Parareal simulation (showing the density map), compared to the ground-truth from the exact simulation. As the algorithm evolves through each cycle, the final state progressively converges to the ground truth. As expected, at the last cycle, the result exactly coincides with the ground-truth simulation. The bottom plot shows the evolution of the SSIM difference between the final state and the ground truth, as a function of Parareal cycles.}}
  \label{FIGURE_parareal_cycles}  
\end{figure}   

Figure-\ref{FIGURE_parareal_cycles} displays the evolution of the last time-step of the last time-window for a Parareal simulation with 40 time-windows, showing how the Parareal result progressively converges to the ground truth. This simulation is run with the electrostatic model, using a neural operator that takes 5 input steps, similar to the surrogates shown in \cite{Carey2024}. At the last cycle, as expected, the Parareal result is identical to the ground truth simulation.

In order to evaluate the efficiency of subsequent Parareal tests, the Structural Similarity Index Measure (SSIM) \cite{Wang2004_SSIM} algorithms is used. As can be observed in Figure-\ref{FIGURE_parareal_cycles}, even though the general structure of the density map at cycles 20 and 30 are similar to the ground truth, the fine details are so different that an MSE comparison between the two would be dominated by point-wise differences rather than inform on the similarity of the blob structures. The SSIM algorithm is designed to provide a better focus on general structures of image maps rather than their exact details. The SSIM is relatively simple to implement in Python and available from the \texttt{skimage.metrics} library. An SSIM value of zero means the two images are completely different, while an SSIM of 1.0 means the two images are identical. The bottom part of Figure-\ref{FIGURE_parareal_cycles} shows the SSIM measure from that Parareal simulation, aligned with the corresponding 2D frames for reference. Note that since the last time-step coincides with the ground truth simulation at the last Parareal cycle, the SSIM will converge to 1.0 at the final cycle. In other words, no matter how bad the coarse-solver is, and how slowly Parareal converges, the SSIM will always go to 1.0 at the final Parareal cycle. Of course, in order to look at very high-precision convergence, an MSE comparison would be preferable, but this is not the objective of this study, and the Parareal algorithm is not aimed at providing extreme precision, but rather reasonable statistical estimation at high speedup.

\begin{figure}[ht!]
  \centering
	\subfigure[ ]{\includegraphics[width=10.0cm]{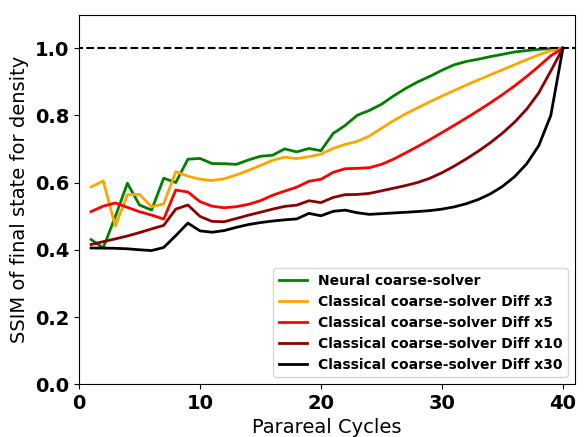}}
  \caption{\scriptsize\textit{\newline Evolution of the SSIM computed for the density map (with the electrostatic model) at the last time-step of the last (10$^{th}$) time-window of a Parareal simulation, compared to the ground-truth from the exact simulation. The evolution is plotted for 5 cases: the Parareal framework with a PDEarena neural operator as the coarse-solver, and 4 `classical' coarse-solvers with increasing levels of additional diffusion. Even for the best `classical' coarse-solver with only 3 times the level of extra diffusion, the SSIM evolution is still outperformed by the Neural-Parareal case.}}
  \label{FIGURE_Neural_vs_Fake}  
\end{figure}

In order to obtain some measure of the efficacy of the Neural-Parareal framework, and the corresponding neural coarse-solver, it is useful to run the same cases with \textit{classical} coarse-solvers. This is achieved by using the actual JOREK code for the coarse solver, with exactly the same physics model, but with a reduced resolution grid of 90$\times$90 instead of 200$\times$200, and increased diffusion coefficients. For this exercise, 4 cases are run with 4 levels of increased diffusion, with all parameters $D$, $\kappa$, $\mu$ increased together by a given factor coefficient $\gamma$. The 4 cases are run with a $\gamma$ of 3, 5, 10 and 30. Effectively, this means that the classical coarse-solver with $\gamma=30$ is a very \textit{bad} or \textit{inaccurate} coarse-solver, while the one with $\gamma=3$ is much more reliable. All cases are run with 40 time-windows, using the electrostatic model, which was the maximum number of time-windows possible for the neural operator which was trained to use 5 time-samples as input. The comparison between these 4 cases and the real neural coarse solver from PDEarena is done by observing the evolution of the SSIM between the last time-step of the last time-window and the ground truth, as a function of Parareal cycles, which is shown in Figure-\ref{FIGURE_Neural_vs_Fake}.

As can be seen in this result, the Neural-Parareal framework with a PDEarena surrogate as coarse-solver outperforms even the best \textit{classical} coarse-solver. It should be noted that, as far as the Parareal method is concerned, such a \textit{classical} coarse-solver with diffusion coefficients increased by only a factor 3, with exactly the same physics model, should be considered an extremely reliable coarse-solver. Effectively, anything better than that should be considered almost identical to the real-solver. Now, evidently, in a non-linear system, any deviation, even at numerical accuracy, can lead to drastic differences for long simulation times, but at least this comparison provides an informative initial measure of how well the neural coarse-solver performs.

Nevertheless, as can be seen from Figure-\ref{FIGURE_Neural_vs_Fake}, after 10/40 cycles (i.e. up to 4$\times$ speedup), even the best Parareal run has barely reached 65\% SSIM accuracy, and by the 20/40 cycle (i.e. up to 2$\times$ speedup), barely 70\%. Although the point of this study is not to investigate the performance of Parareal as a speed-up option for the JOREK code, a relevant point is to investigate how far the neural operators can be improved to increase the precision (and/or speed-up) of the Parareal framework.

\subsection{Integrated framework demonstration}

As mentioned earlier, for a specific number $n$ of time-windows, assuming the i/o and orchestration operations are negligible compared to the fine-solver, the Parareal algorithm will provide a real-time speed-up of $\frac{n}{i}$, where $i$ is the number of Parareal cycles needed to reach the desired level of convergence (the exact speed-up of $\frac{n}{i}$ might be reduced depending on the cost of the coarse-solver, the i/o operations and the orchestration of all jobs). However, assuming $n$ is large, then the total amount of fine-solver time-steps simulated will be $i$ times the amount that a single (non-Parareal) simulation would provide. Given this large amount of data produced by the Parareal framework, a sensible approach is to consider a situation where a scientist would need to produce new simulations progressively within an input-parameter domain to explore new features of the problem at hand. Initially, since no data is available, it is impossible to train any neural operator, but after a reasonable number of simulations are achieved, a first (very) coarse neural solver can be trained, thus enabling simulations to be run with the Parareal framework. From that point onward, all simulations run with Parareal produce more data, which can then be bootstrapped into the training of progressively more accurate neural solvers. Ideally, the more simulations are run, the more training data, the better the neural solver, and thus the higher Parareal speed-up. In the ideal scenario that the neural solver can become as accurate as the fine-solver itself, this means simulations could then reach a speed-up with Parareal equal to the number of time-windows, i.e. with $n=100$, simulations would be achieved 100 times faster.

This optimistic vision will be highly dependent on the use case. For simple problems where neural solvers can indeed reach high accuracy levels, such speed-up results might be achievable, but one would argue that simple sets of PDEs do not need to be parallelised in the first place, as they would be easily and quickly evaluated with ordinary numerical solver methods. A demonstration of this bootstrap method is provided here with the electromagnetic model of Reduced-MHD. Since a lot of data was already available for the electrostatic model, it was intriguing to explore an entirely new case where no data was available, in order to provide a genuine test of the idea of Neural-Parareal.

First, a set of 20 simulations is run, without Parareal. Each one with a random sampling of the initial conditions of multiple blobs. Once these 20 simulations are obtained, the time-trajectories of these are each split into segments of the size of the time-windows. In order to enable enough data within a time-window to be used for the training of the neural solver, the time-domain is decomposed into 20 time-windows. This results in segments of data containing 100 time-steps of simulations, meaning 10 data frames (due to down-sampling of the frequency agreed for the neural solver input). The neural solver itself is trained to use 5 time-samples as inputs (to output a single one, which can be rolled-out). 

After the first neural coarse-solver is trained, another set of 20 simulations is run with Parareal using the neural coarse-solver, starting from a new set of random inputs. Each simulation is run for the full 20 Parareal cycles. Thus the total number of time-windows run for each Parareal simulation is $\sum_{i=1}^{n}i$ = $\frac{n(n+1)}{2}$, which gives 210 with $n=20$. This new batch of 20 simulations produces a total of 4200 fine-solver segments that can be added to the initial data set to re-train the neural coarse-solver. 

\begin{figure}[ht!]
  \centering
	\subfigure[ ]{\includegraphics[width=8.0cm]{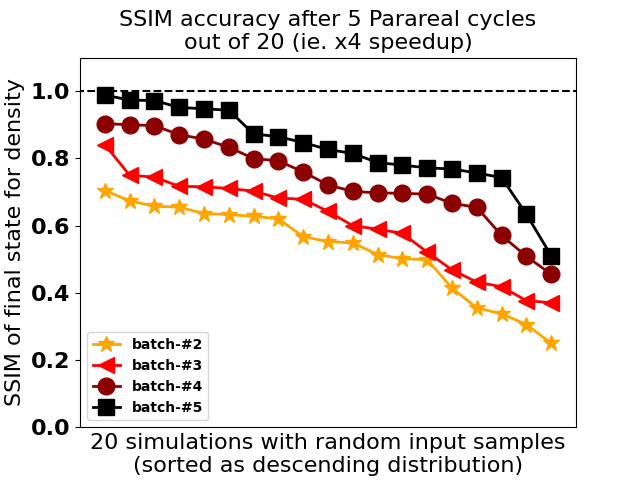}}
	\subfigure[ ]{\includegraphics[width=8.0cm]{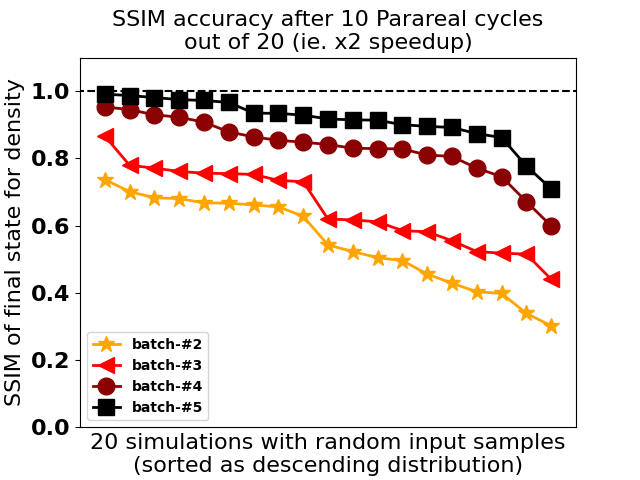}}
  \caption{\scriptsize\textit{\newline Performance of the Neural-Parareal framework evaluated at \textnormal{(a)} cycle 5 out of 20 (i.e. speed-up of 4), and at \textnormal{(b)} cycle 10 out of 20 (i.e. speed-up of 2). For each batch (different colors/signs), all simulations are ordered in descending order to provide a distribution-like view of the performance.)}}
  \label{FIGURE_Main_Result}  
\end{figure}   

With this new coarse-solver, another set of 20 simulations is run with Parareal, and the resulting data aggregated to the existing dataset for a new training, and so on. This Neural-Parareal loop, described in Figure-\ref{FIGURE_Neural_Parareal} is run for 4 iterations, and the result is shown in Figure-\ref{FIGURE_Main_Result}, demonstrating clear performance improvement at each iteration of the framework. The framework implementation, although specific to the JOREK code for now, is available on \cite{github_neural_parareal}.

Although in a normal situation one may wish to stop after a fixed number of parareal cycles if convergence is not achieved, for the purpose of this exercise, running all the Parareal cycles ensures more data is created for future training, and it also enables a direct comparison against the ground truth, which would not be available otherwise. All final simulation trajectories (without the additional Parareal intermediate windows) can be downloaded from Zenodo \cite{Zenodo_dataset_RMHD}.

\begin{figure}[ht!]
  \centering
	\subfigure[ ]{\includegraphics[width=8.0cm]{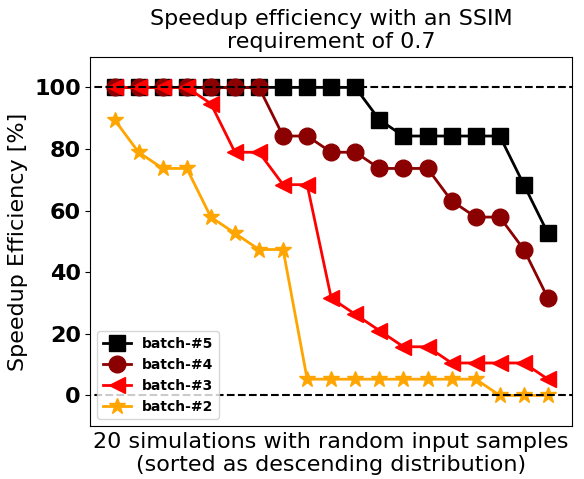}}
	\subfigure[ ]{\includegraphics[width=8.0cm]{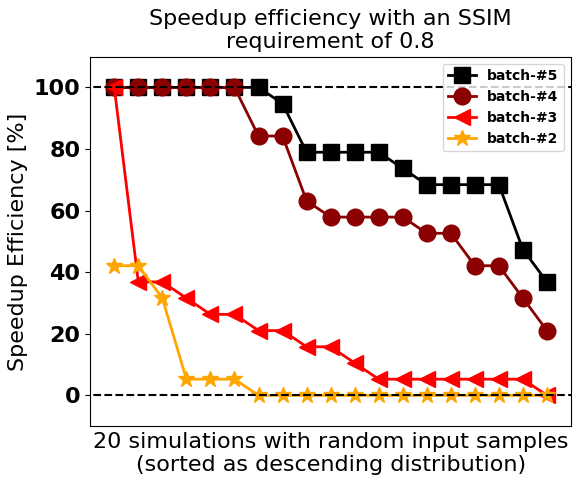}}
  \caption{\scriptsize\textit{\newline Speed-up efficiency of the Neural-Parareal framework for a required SSIM-accuracy of \textnormal{(a)} 70\%, and \textnormal{(b)} 80\%. For each batch (different colors/signs), all simulations are ordered in descending order to provide a distribution-like view of the speed-up. In this particular example, since the Parareal simulations have 20 time-windows, a 100\% speed-up efficiency means a time-to-solution accelerated by a factor 20 (for the required SSIM-accuracy), whereas a speed-up efficiency of 50\% means an acceleration of a factor 10.)}}
  \label{FIGURE_Main_Result2}  
\end{figure}

The result in Figure-\ref{FIGURE_Main_Result} demonstrates that in a Neural-Parareal framework, the more simulations are run, the more precise the neural coarse-solver becomes. As the neural coarse-solver improves, the speed-up obtained by the Parareal framework will increase. This is best illustrated in Figure-\ref{FIGURE_Main_Result2}, by plotting the speed-up efficiency of each Parareal simulation for a given SSIM accuracy requirement. The maximal speedup of a Parareal simulation is the number of time-windows (assuming the number of computing resources are scaled linearly with the number of time-windows). If it takes a single evaluation of the coarse-solver (and fine-solver) on each time-window to obtain the required accuracy level, then the speed-up (time-to-solution acceleration) is equivalent to the number of time-windows. In other words, the speed-up efficiency is 100\%. With every Parareal iteration, the speed-up efficiency diminishes. As shown in Figure-\ref{FIGURE_Main_Result2}, the speed-up efficiency increases significantly for each batch of simulation, when the neural operator coarse-solver is updated.

\section{Conclusion}\label{SECTION_CONCLUSION}

\subsection{Summary}

This paper presents the development of an integrated Neural-Parareal framework, which bootstraps the training of neural coarse-solvers as more simulations are being produced by users, progressively leading to more and more accurate neural operators, and thus larger speed-ups of the Parareal simulations. The framework exploits the large amount of data that Parareal frameworks produce by design. The Parareal algorithm can provide real-time speedup (i.e. time-to-solution) of simulations for a given precision requirement. Provided a very fast and precise coarse-solver is available, and given large amounts of High-Performance-Computing resources if the simulation can be split into $n$ time-windows on $n$ parallel computing resources, the speedup can potentially become $n/2$ or $n/3$. If $n$ is large, of the order of hundreds, this speedup can be significant. This trade of computing cost versus time-to-solution also comes with increased simulation data production, which is ideal for training deep learning models and, in this context, neural operators.

The demonstration of this self-improving framework is demonstrated using MHD simulations relevant to fusion research, with radially evolving blobs in a 2D poloidal slab with toroidally axisymmetric. The neural operators are trained using PDEarena, and the MHD simulations are performed with the JOREK code. Parallelisation of the framework is implemented with SLURM on HPC clusters. For a set of Reduced-MHD equations, the framework is evolved for 4 iterations using 20 new simulations at each iteration, clearly showing rapidly improving performance.

Beyond this demonstration, several improvements could be implemented. Particularly in the urgent context of Digital Twins for fusion, this self-improving platform could take full advantage of the rapidly evolving domain of Artificial Intelligence and exascale computing, revealing an interesting avenue in the convergence of AI and HPC.

\subsection{Potential extensions and improvements}

In order to build upon the work presented here, the following ideas could be addressed in the near future:
\begin{enumerate}
    \item \textbf{Using a real Parareal convergence measure} \newline In the demonstration above, each Parareal simulation is run for the full number of iterations possible. This is partly to create more data for the training, but also to provide a clean performance measure by comparing against the real simulation. In the future, instead of using SSIM (or MSE) against the ground truth of the final simulation result, one could compare each Parareal iteration against the previous iteration to check the relative convergence. How to properly measure Parareal convergence is a question in itself. A preview of this is given in Figure-\ref{FIGURE_relative_convergence}, with the SSIM accuracy compared to the previous Parareal iteration (instead of the ground truth) for a speed-up of 2 and a speed-up of 4.
\begin{figure}[ht!]
  \centering
	\subfigure[ ]{\includegraphics[width=8.0cm]{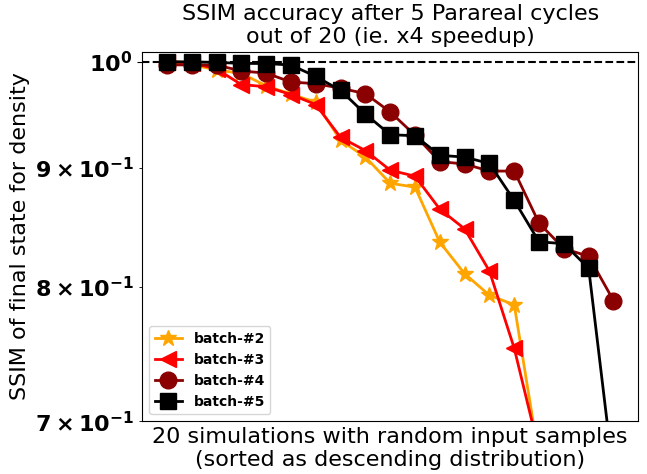}}
	\subfigure[ ]{\includegraphics[width=8.0cm]{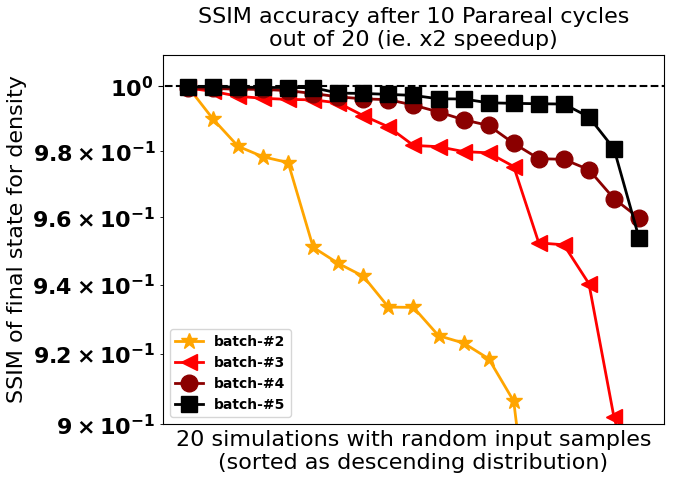}}
  \caption{\scriptsize\textit{\newline Performance of the Neural-Parareal framework evaluated at \textnormal{(a)} cycle 5 out of 20 (i.e. speed-up of 4), and at \textnormal{(b)} cycle 10 out of 20 (i.e. speed-up of 2). Here the SSIM is evaluated against the final time-step of the previous Parareal iteration, as opposed to the ground truth (assuming the ground truth is not available).)}}
  \label{FIGURE_relative_convergence}  
\end{figure}   
    \item \textbf{Higher resolution surrogates} \newline As mentioned in Section-\ref{SECTION_NN}, the spatial/temporal resolution of the neural operator is down-sampled from the simulation. Although the full resolution equivalent to the simulation may not be required, it will undoubtedly affect the precision of the Parareal framework as a whole, and should therefore be investigated.
    \item \textbf{Other PDEarena options} \newline The high performance of the FNO option in PDEarena may be dependent on the use case, as addressed in Section-\ref{SECTION_NN}, and it would be interesting to investigate other options, ideally using multiple options at once and choosing the optimal one as part of a broader hyperparameter tuning.
    \item \textbf{Using existing foundation models} \newline Instead of training neural operators from scratch, one could directly fine-tune existing foundation models to create the first coarse-solvers of a given simulation set-up if the amount of data is sparse. This would be particularly relevant to realistic use cases, where engineers and researchers are regularly faced with new problems, including new geometries, new meshes and new physics models.
    \item \textbf{Filtering and discarding data} \newline A particularly attractive aspect of the above point is that fine-tuning of large models would be more suited to the potential requirement of discarding data after simulations have been run, with respect to available data-storage capabilities. Regardless, even with the current framework, one could retain only the data of simulations that have struggled to converge, and discard data from simulations that have converged quickly, and thus are already covered by the surrogate.
    \item \textbf{Better input-domain sampling with Active Learning} \newline For a self-improving framework like the one presented here, it may be more efficient to sample the input-domain space with advanced methods like Active Learning, rather than letting users choose new simulations at random. In other words, a modern framework would ingest new input requests from users, and provide a first estimation of whether the new simulation is actually needed, or whether several simulations have already been performed in the vicinity, thus implying the coarse neural solver is already reliable, and requesting confirmation before this specific simulation is run.
\end{enumerate}

\newpage

\section*{Acknowledgement}

This work was performed with the support of the JOREK Team [See https://www.jorek.eu for the present list of team members].

This work has been carried out within the framework of the EUROfusion Consortium, funded by the European Union via the Euratom Research and Training Programme (Grant Agreement No 101052200 — EUROfusion). Views and opinions expressed are however those of the author(s) only and do not necessarily reflect those of the European Union or the European Commission. Neither the European Union nor the European Commission can be held responsible for them. 

This work has been carried out within the framework of the RCUK Energy Programme [grant number EP/I501045].

This work was performed using the MARCONI computer at CINECA in Italy, within the EUROfusion framework.

This work was performed using the Cambridge Service for Data Driven Discovery (CSD3), part of which is operated by the University of Cambridge Research Computing on behalf of the STFC DiRAC HPC Facility (www.dirac.ac.uk). The DiRAC component of CSD3 was funded by BEIS capital funding via STFC capital grants ST/P002307/1 and ST/R002452/1 and STFC operations grant ST/R00689X/1. DiRAC is part of the National e-Infrastructure.

\newpage

\bibliography{my_references}

\end{document}